\documentclass[letterpaper,twocolumn]{esapub} 
\usepackage{cranm_timesv}
\usepackage{epsfig}
\usepackage{natbib}

\def\gtrsim{\mathrel{\hbox{\rlap{\hbox{\lower4pt\hbox{$\sim$}}}\hbox{$>$}}}}
\def\lesssim{\mathrel{\hbox{\rlap{\hbox{\lower4pt\hbox{$\sim$}}}\hbox{$<$}}}}

\title{\large
Capabilities of UV Coronagraphic Spectroscopy for Studying the
Source Regions of Solar Energetic Particles and the Solar Wind}

\author{John L. Kohl}
\author{Steven R. Cranmer}
\author{Larry D. Gardner}
\author{Jun Lin}
\author{John C. Raymond}
\author{Leonard Strachan}
\affil{Harvard-Smithsonian Center for Astrophysics,
Cambridge, MA 02138, USA}

\begin{document}


\maketitle

\begin{abstract}
We summarize the unique capabilities of UV coronagraphic spectroscopy
for determining the detailed plasma properties (e.g., density,
temperature, outflow speed, composition) of the source regions of
both transient phenomena such as CMEs, flares, and solar energetic
particles (SEPs) and more time-steady solar wind streams.
UVCS/SOHO observations have provided the first detailed diagnostics
of the physical conditions of CME plasma in the extended corona.
It provided new insights into the roles of shock waves, reconnection,
and magnetic helicity in CME eruptions.
We summarize past observations and discuss the diagnostic
potential of UV coronagraphic spectroscopy for characterizing two
possible sites of SEP production: CME shocks and reconnection
current sheets.  UVCS/SOHO has also led to fundamentally new views of
the acceleration region of the solar wind.  Understanding the physical
processes in this region, which ranges from the low corona
($r = 1.1$--1.5 $R_{\odot}$) past the sonic points
($r > 5 \, R_{\odot}$), is key to linking the results of solar
imaging to {\em in situ} particle and field detection.
Despite the advances that have resulted from UVCS/SOHO, more
advanced instrumentation could determine properties of additional
ions with a wider sampling of charge/mass combinations.
This would provide much better constraints on the specific kinds
of waves that are present as well as the specific collisionless
damping modes.
Electron temperatures and departures from Maxwellian velocity
distributions could also be measured.
The instrumentation capable of making the above observations
will be described.
\end{abstract}

\vspace*{-8.05in}
\noindent
{\small
To be published in the proceedings of
{\em Solar Wind 11/SOHO--16: Connecting Sun and Heliosphere,}
June 13--17, 2005, Whistler, Canada,
ESA SP--592.}

\vspace*{7.33in}
\section{Introduction}

In the following, we describe the capabilities of ultraviolet
coronagraphic spectroscopy to address the SEP/flare/CME and solar
wind problems, and we provide a brief
description of a science payload that is
capable of carrying out the required observations.

\section{CMEs, Flares, and SEPs}

UVCS/SOHO observations have
provided new insights into the roles of shock waves,
reconnection, and magnetic helicity in CME eruptions
(Raymond 2002).

\subsection{CME Shocks}

The key parameters in theories of particle acceleration by
shocks are the pre-shock plasma conditions (including seed particle
population), the shock speed, and the angle between the magnetic field
and the shock motion.

UVCS has observed CME-driven shocks through their effect on the
widths of UV spectral lines (Raymond et al.\  2000;
Mancuso et al.\  2002).
These observations provide information about the compression ratio in
the shock, a crucial parameter for predicting SEP spectra, and
information about the thermal equilibration among electrons, protons
and heavier ions. Electron heating is relatively modest, and the line
widths of oxygen and silicon ions imply temperatures far higher than
the proton temperatures, a potentially important consideration for
models of SEP composition.
In some events, the shock compression ratio can be determined
from Type II burst band splitting (e.g., Vr\v{s}nak et al.\  2002),
but not all radio bursts contain enough detail for this diagnostic
to be useful.
For a much larger fraction of events, UV spectroscopy can be used
to determine the compression ratio via two independent techniques:
(1) measuring pre- to post-shock temperature ratios (i.e.,
both $T_p$ and $T_e$) with resonant and Thomson-scattered
H~I Ly$\alpha$ line widths, then using adiabatic theory to compute
the density ratio;
(2) measuring ion temperatures of several species having different
charges and masses and applying collisionless theories of multi-ion
shock heating (e.g., Lee et al.\  1987; Lee \& Wu 2000) to
compute the compression ratio that is most consistent with the
observations (for both methods, see Mancuso et al.\  2002).

UVCS routinely obtains the densities, ionization states and elemental
abundances in the pre-CME corona (e.g., Raymond et al.\  2003). The
densities obtained by UVCS can be combined with Type II radio burst
drift rates to obtain shock speeds. Upper limits on the coronal
Alfv\'{e}n speed above active regions were inferred from the
derived shock speeds by requiring that the disturbances propagate at
least as fast as the local characteristic speed (Mancuso et al.\  2003).
For a subset of
events, the resulting shock speeds are in much better agreement with
LASCO CME expansion rates than shock speeds based upon average coronal
density profiles, although there are some uncertainties in the
measured drift rates (Mancuso \& Raymond 2004).
The Alfv\'{e}n and shock speeds can be inferred from detection of the
shock arrival at different heights as determined by the timing
of the increase in line widths of UV emission lines (e.g.,
Ciaravella et al.\  2005).
The angle between the shock front and the magnetic field
requires the pre-shock field direction, which can be determined from
streamer morphology. Severe elemental depletions are often observed in
the closed field portions of streamers (Uzzo et al.\  2004), providing
an additional indicator of field topology.
Another parameter potentially vital to the efficiency of shock
acceleration is the density of suprathermal seed particles (e.g.,
Desai et al.\  2003). While UVCS was not able to detect such particles,
the improved sensitivity and instrumental profile characterization
of next-generation instruments will make it possible to
determine suprathermal particle densities out to $\sim$6 times the
mean proton thermal speed (Cranmer 1998).

Although UVCS has proven the feasibility of detecting and characterizing
CME shocks for several representative events, next-generation
instrumentation can provide more extensive diagnostics and more
complete spatial and temporal coverage.
The empirical characterization of the coronal shock conditions and
the ambient solar wind properties can then be used as inputs to:
(1) 3D MHD models of the shock propagation through the heliosphere,
and (2) multi-scale models of the SEP acceleration, transport,
and energy spectrum synthesis
(Zank et al.\  2000; Li et al.\  2003; Rice et al.\  2003).
Such constraints need to be applied in order to model specific
events, such as the massive solar storms of Oct--Nov 2003.
SEP acceleration and transport models can be tested for specific
events by using spectroscopy and other remote-sensing diagnostics to
constrain the initial parameters of the shock in the corona,
then comparing with observed SEP energy spectra from, e.g., the
Inner Heliospheric Sentinels.
Iterative testing and refinement will ultimately result
in a comprehensive validation of a predictive SEP model.

\subsection{CME Current Sheets}

Models of CMEs rely heavily on reconnection in current sheets,
either trailing beneath the ejected magnetic flux rope or
creating the flux rope in the first place (see Klimchuk 2001).
Reconnection dumps large 
amounts of energy in the lower atmosphere of the Sun, creating intense 
heating, which accounts for the traditional flare ribbons and loops 
and for the current sheet containing hot plasma (Forbes et al.\  1989; 
\v{S}vestka \& Cliver 1992; Forbes \& Acton 1996; Priest \& Forbes 2002).

UVCS observations made it possible for the first time to allow us 
to carry out diagnostics of the plasma inside the current sheet 
(Ciaravella et al.\  2000; Ko et al.\  2003).
A narrow feature was seen in [Fe~XVIII] emission in the space
between the post-flare loops and the CME core, indicating
electron temperatures near $6\times 10^{6}$ K.
Significant progress in studying the current sheet and the process of 
magnetic reconnection in the current sheet was made recently when the UV 
spectral data of the plasma inside the current sheet during two events on 
January 8, 2002 and on November 18, 2003 were obtained and analyzed
(Ko et al.\  2003; Lin et al.\  2005).  Both events developed a fast 
CME, a growing flare loop system, and a long current sheet that connects 
CME and flare loops.
In one of these events, the pattern of the reconnection inflow near the 
current sheet was well recorded in H~I Ly$\alpha$.
This allowed us to deduce the speed of the reconnection 
inflow directly and even to estimate the thickness of the current sheet.
Also, it is possible to determine the magnetic field strength
by using the observed speeds, densities, and temperatures to compute
the kinetic and thermal energy densities in the reconnection region,
then assume that this energy comes from the annihilation of an
equal amount of magnetic energy just outside the current sheet
(Ko et al.\  2003).
All of the parameters described above are required in order to put
empirical constraints on the {\em reconnection rate} and electric field
strength in the current sheet.

Such progress has very important theoretical consequences:
for example, we are able 
to deduce the electrical resistivity (conductivity) in the current sheet, 
or in the reconnection region.
Results obtained from UVCS and
other remote-sensing instruments can provide the value of the 
electrical conductivity of the plasma inside the current sheet in an
ongoing eruption for the first time
since the impetus of applying reconnection theory to 
solar eruptions began six decades ago (Giovanelli 1946; and also see 
Priest \& Forbes 2000).

With the knowledge of the dynamical process inside the current sheet,
we are further able to investigate the particle
acceleration taking place in the current sheet.
A strong electric field is induced by magnetic reconnection 
in the current sheet. For a typical event, the electric field strength
reaches about 5 V/cm (e.g., Wang et al.\  2003;
Qiu et al.\  2004 for observations, and
Martens \& Kuin 1989; Forbes \& Lin 2000; Lin 2002 for theories).
An extremely high value of 50 V/cm was 
also reported (Xu et al.\  2004). In principle, such a strong electric 
field is able to accelerate any charged particles.
The current sheet is an assembly of waves and electric field, and 
accelerations can occur in various ways (see also
Miller \& Roberts 1995; Litvinenko 2000).
We do not yet know exactly what 
happens in a real eruptive process, but the observations that a series of 
bright blobs flow successively out of the current sheet (e.g.,
Ko et al.\  2003; Lin et al. 2005)
replicate one of the main characteristics of 
magnetic reconnection inside the turbulent current sheet: the turbulent 
eddies or small magnetic islands inside the current sheet tend to merge 
into bigger ones before they leave the current sheet
(Ambrosiano et al.\  1988).

Further progress in understanding the above processes occurring in 
the current sheet depends on the accurate measurement of the 
thickness of the current sheet,
plasma parameters in the current sheet (including electron and ion
velocity distributions and densities),
the speeds of reconnection inflow/outflow near the current sheet,
as well as electric and magnetic fields in and around the current sheet.
UV coronagraphic spectroscopy is uniquely suited to these requirements.

\section{The Solar Wind}

UVCS has led to fundamentally new views of the
{\em acceleration region} of the solar wind.
By measuring emission lines formed both by collisional excitation
and by the resonant scattering of solar-disk photons, UV
spectroscopy provides a multi-faceted characterization of the
kinetic properties of atoms, ions, and electrons (e.g.,
Withbroe et al.\  1982; Cranmer 2002a).
The Doppler-broadened shapes of emission lines are direct probes
of line-of-sight (LOS) particle velocity distributions (i.e.,
essentially providing $T_{\perp}$ when the off-limb magnetic field
is $\sim$radial), and red/blue Doppler shifts reveal bulk flows
along the LOS.
Integrated intensities of resonantly scattered lines can be
used to constrain the solar wind velocity and other details
about the velocity distribution in the radial direction
(e.g., $u_{\parallel}$ and $T_{\parallel}$); this is the so-called
``Doppler dimming/pumping'' diagnostic (e.g., Noci et al.\  1987).
Intensities of collisionally dominated lines---especially when
combined into an emission measure distribution---can
constrain electron temperatures, densities, and elemental
abundances in the coronal plasma.
Even departures from Maxwellian and bi-Maxwellian velocity
distributions are detectable with spectroscopic measurements
having sufficient sensitivity and spectral resolution
(e.g., Cranmer 2001).

In the high-speed solar wind, UVCS measured outflow speeds
that were found to become supersonic much closer to the Sun
than previously believed.
In coronal holes, heavy ions (e.g., O$^{+5}$) were found to
flow faster, to be heated hundreds of times more strongly than
protons and electrons, and to have anisotropic temperatures with
$T_{\perp} > T_{\parallel}$ (Kohl et al.\  1997, 1998, 1999;
Cranmer et al.\  1999).
These unexpected results have rekindled theoretical efforts to
understand the heating and acceleration of the fast wind in the
extended corona (e.g., Tu \& Marsch 1997;
Leer et al.\  1998; Axford et al.\  1999; Hollweg 1999;
Hollweg \& Isenberg 2002; Marsch 2004).

The slow solar wind was found to flow mostly along the
outer edges of bright streamers, near locations with measured
abundance patterns matching those of the {\em in situ} slow wind
(Strachan et al.\  2002; Raymond et al.\  1997).
The closed-field ``core'' regions of streamers, though, exhibit
heavy element abundances only 3 to 30\% of those seen at 1 AU,
indicating gravitational settling (e.g., Raymond 1999;
V\'{a}squez \& Raymond 2005).
UVCS observed the transition from a high-density
collision-dominated plasma at low heights in streamers to a
low-density collisionless plasma at large heights, the latter
exhibiting high ion temperatures and anisotropies that suggest similar
physics as in the fast wind (Ko et al.\  2002; Frazin et al.\  2003).

If the kinetic properties of {\em additional ions} were to be measured
in the extended corona (i.e., a wider sampling of charge/mass
combinations) we could much better constrain the specific kinds of
waves that are present as well as the specific collisionless damping
modes (e.g., Cranmer 2002b).
Measuring the {\em electron temperature} above $\sim$1.5 $R_{\odot}$
(never done directly before) would finally allow us to determine
the bulk-plasma heating rate in different solar wind structures,
thus putting the firmest ever constraints on models of why the
slow [fast] speed wind is slow [fast] (e.g.,
Suess et al.\  1999; Endeve et al.\  2004; Cranmer, these
proceedings).
Measuring {\em non-Maxwellian velocity distributions} of
electrons and positive ions would allow us to test specific models
of MHD turbulence, cyclotron resonance, and velocity filtration.
New capabilities such as these would be enabled by greater photon
sensitivity, an expanded wavelength range, and the use of measurements
that heretofore have only been utilized in a testing capacity
(e.g., Thomson-scattered H~I Ly$\alpha$ to obtain $T_e$;
the Hanle effect to obtain constraints on the magnetic field).
These would then allow
the relative contributions of different physical processes
to the heating and acceleration of all solar wind plasma
components to be determined directly.

\begin{figure}
\centering
\epsfig{figure=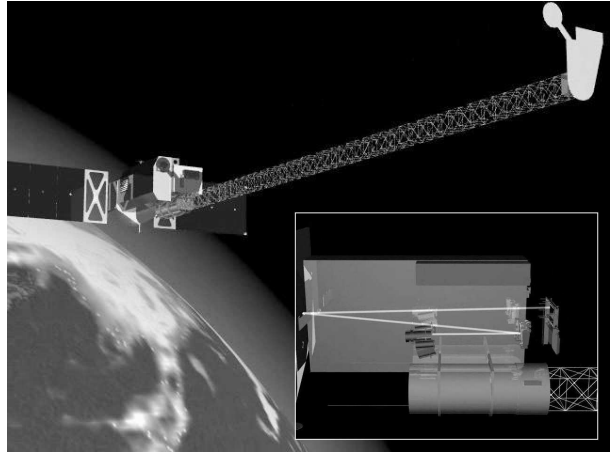,width=8.0cm}
\caption{%
Spacecraft concept for
advanced large-aperture coronagraphs.
Inset shows a
diagram of an Advanced UV Coronagraph Spectrometer.}
\end{figure}

\section{Implementation}

A mission capable of carrying out the required observations would
include two instrument units: a large-aperture
ultraviolet coronagraph spectrometer (see Fig.\  1) and a
large-aperture visible light coronagraph.
A suitable design was developed during a MIDEX Feasibility Study
for a mission called ASCE.
Remote external
\linebreak[4]
occulters 
supported by a deployable boom provide much larger unvignetted apertures 
than are possible with conventional coronagraphs. These instruments 
provide major improvements in sensitivity, stray light rejection, 
spatial resolution, minimum observable height and ultraviolet wavelength 
range. New spectroscopic diagnostics for the electron velocity 
distribution, magnetic field, and parameters for a broad range of newly 
observable ions are implemented by these instruments. Unprecedented 
cadences are possible.


This work is supported by NASA
under grants
NNG\-04G\-E77G, and
NNG\-04G\-E84G to the Smithsonian Astrophysical Observatory.

\end{document}